\documentclass[fleqn,usenatbib]{mnras}

\usepackage{newtxtext,newtxmath}

\usepackage[T1]{fontenc}
\usepackage{booktabs}
\usepackage{longtable} 
\usepackage{lipsum} 
\usepackage{array} 
\usepackage{natbib}
\newcommand{\citeay}[1]{\citeauthor{#1}, \citeyear{#1}}
\DeclareRobustCommand{\VAN}[3]{#2}
\let\VANthebibliography\thebibliography
\def\thebibliography{\DeclareRobustCommand{\VAN}[3]{##3}\VANthebibliography}

\usepackage{graphicx}	
\usepackage{amsmath}	

\title[LGRB Progenitor]{Pieces of evidence for multiple progenitors of Swift long gamma-ray bursts}

\author[Y. K. Qu et al. ]{
Yan-Kun Qu,$^{1}$
Zhong-Xiao Man,$^{1}$
and Zhi-Bin Zhang$^{1}$\thanks{E-mail: z\_b\_zhang@sina.com (ZHB)}
\\
$^{1}$School of Physics and Physical Engineering, Qufu Normal University, Qufu, 273165, People's Republic of China\\
}

\date{Accepted XXX. Received YYY; in original form ZZZ}

\pubyear{\the\year{}}

\begin{document}
\label{firstpage}
\pagerange{\pageref{firstpage}--\pageref{lastpage}}
\maketitle

\begin{abstract}
Long gamma-ray bursts~(LGRBs) are typically thought to result from the collapse of massive stars. Nonetheless, recent observations of gamma-ray bursts ~(GRBs)  211211A and  230307A, coupled with the low-redshift excess of LGRB event rates relative to star formation rates, present significant challenges to the prevailing model. We reexamine the selection criteria for higher redshift complete GRB samples and identify 280 Swift GRBs with peak flux over \(2.6 \, \text{ph} \, \text{cm}^{-2} \, \text{s}^{-1}\). Assuming all LGRBs with \(z \geq 2\) originate from collapsars, we construct the GRB luminosity functions~(LFs) in three scenarios: no evolution, luminosity evolution, and density evolution. Our results indicate that a strong redshift evolution in luminosity~(\(\delta = 1.87^{+0.27}_{-0.31}\)) or in density~(\(\delta = 1.10^{+0.21}_{-0.20}\)) is necessary. The luminosity/density evolution model predicts 72.67/57.28 collapsar GRBs at \(z < 2\), which can account for 67.29\%~/ ~53.04\% of the observed LGRBs. 
This suggests that a substantial portion of LGRBs at $z< 2$ may not be collapsar GRBs, which would challenge the universality of empirical GRB relations and affect their reliability in cosmological applications.
\end{abstract}

\begin{keywords}
Gamma-ray bursts -- Luminosity function -- Star formation
\end{keywords}



\section{Introduction}
Gamma-ray bursts~(GRBs) are highly energetic and intense explosions of $\gamma$-ray radiation on a short timescale. These events exhibit a diverse array of spectral and temporal characteristics, making them a subject of significant interest in astrophysical research. Since 2004, Swift satellite has detected more than 1,600 GRBs, about 500 out of which have the measured redshifts. Specifically, GRBs are traditionally classified into two categories, that is Long gamma-ray bursts~(LGRBs) and short gamma-ray bursts~(SGRBs) divided by a duration time of $T_{90}=2$ s. The  criterion was  verified  by  KONUS-Wind~\citep{1981Ap&SS..80....3M},  BATSE~\citep{1993ApJ...413L.101K} and Swift satellites~\citep{2008A&A...484..293Z}.
LGRBs are generally believed to originate from the collapse of massive stars~\citep{1993ApJ...405..273W,1998ApJ...494L..45P}, as they are often found in star-forming regions~\citep{1997ApJ...486L..71T,1998ApJ...507L..25B}, and a significant fraction of them are found to associate  with supernovae~\citep{1998Natur.395..670G,2008ApJ...687.1201K,2010MNRAS.405...57S}. Consequently, the event rate of LGRBs  is expected to correlate with the star formation rate~(SFR)~\citep{2012A&A...539A.113E,2013ApJ...772...42H,2016A&A...587A..40P}. In contrast, SGRBs are believed to originate from  mergers of compact binary systems~\citep{2014ARA&A..52...43B}, as evidenced by their occurrence in regions distant from star-forming areas~\citep{2013ApJ...769...56F} and associations with gravitational wave events~\citep{2017PhRvL.119p1101A,2017ApJ...848L..14G} or kilonovae~\citep{2013Natur.500..547T,2019MNRAS.489.2104T,2021ApJ...906..127F},~suggesting that the event rate of SGRBs  may be connected with  a delayed SFR~\citep{2018MNRAS.473.3385P,2024ApJ...960...77D}.

The origin of LGRBs has recently been challenged. 
For most astronomical sources, a typical forms of power-law~(PL) or broken power-law~(BPL) is usually adopted to describe the  luminosity function~(LF)~(For a more detailed discussion, please refer to Section ~\ref{sec:cite}). However, a triple form of the LF is found to exist in  LGRBs~\citep{2015ApJ...812...33S,2021MNRAS.508...52L}. The  triple power-law~(TPL) form manifests  that the classes of LGRBs may be more complex than before. Recently,~\cite{2023ApJ...958...37D}  found that the event rate of high-luminosity LGRBs follows the SFR, while the  low-luminosity LGRBs exhibits higher event rate than the SFR at the lower redshifts, which  indicates that high- and low-luminosity LGRBs  should have distinct progenitors.~\cite{2009ApJ...703.1696Z} found that three LGRBs~(050724, 060614 and 061006) exhibit typical characteristics of merger GRBs. In addition,  GRB 211211A~\citep{2022Natur.612..223R} as a LGRB with \(T_{90} > 30s\)  and  \(z = 0.076\)  was found to  associate with a kilonova  instead of a traditional supernova. Similarly, GRB 230307A ~\citep{2024Natur.626..737L} with  \(z = 0.0646\) and \(T_{90} \sim 35s\) was identified as abnormal product of  a binary neutron star merger. According to the comparison of the event rate of LGRBs with the SFR~\citep{2024ApJ...963L..12P},  it is suggested that more than 60\% of LGRBs at \(z < 2\) may be non-collapsars.

In this letter, we set the selection criteria to derive an updated complete GRB sample and  study the composition of low-redshift LGRBs. In Section~\ref{sec:Samples and Methods}, we elaborate the sample  selection and construct their LF in Section~\ref{sec:ANALYSIS METHOD}. Section~\ref{sec:RESULTS} presents our results, and the final section provides a concise discussion of our findings.

\section{Samples}
 \label{sec:Samples and Methods}
 Since 2004, the Swift satellite has detected more than 1,600 GRBs, among which approximately 500 have measured redshifts. To construct a complete  LGRB~($T_{90}>2s$) sample following the criteria outlined by~\cite{2006A&A...447..897J},~\cite{2012ApJ...749...68S} summarized five key selection criteria:
(1) The burst must be accurately localized by Swift/XRT, with its coordinates rapidly disseminated; (2) The Galactic extinction along the line of sight to the burst should be low ~(\(A_V < 0.5\)); (3) The declination of the GRB must fall within the range \( -70^\circ < \delta < 70^\circ \); (4) The angle between the Sun and the field of the burst must exceed 55° ~(\(\theta_{\text{Sun}} > 55^\circ\)); and (5) There should be no nearby bright stars that could interfere with observations.

\begin{figure}
\includegraphics[width=\columnwidth]{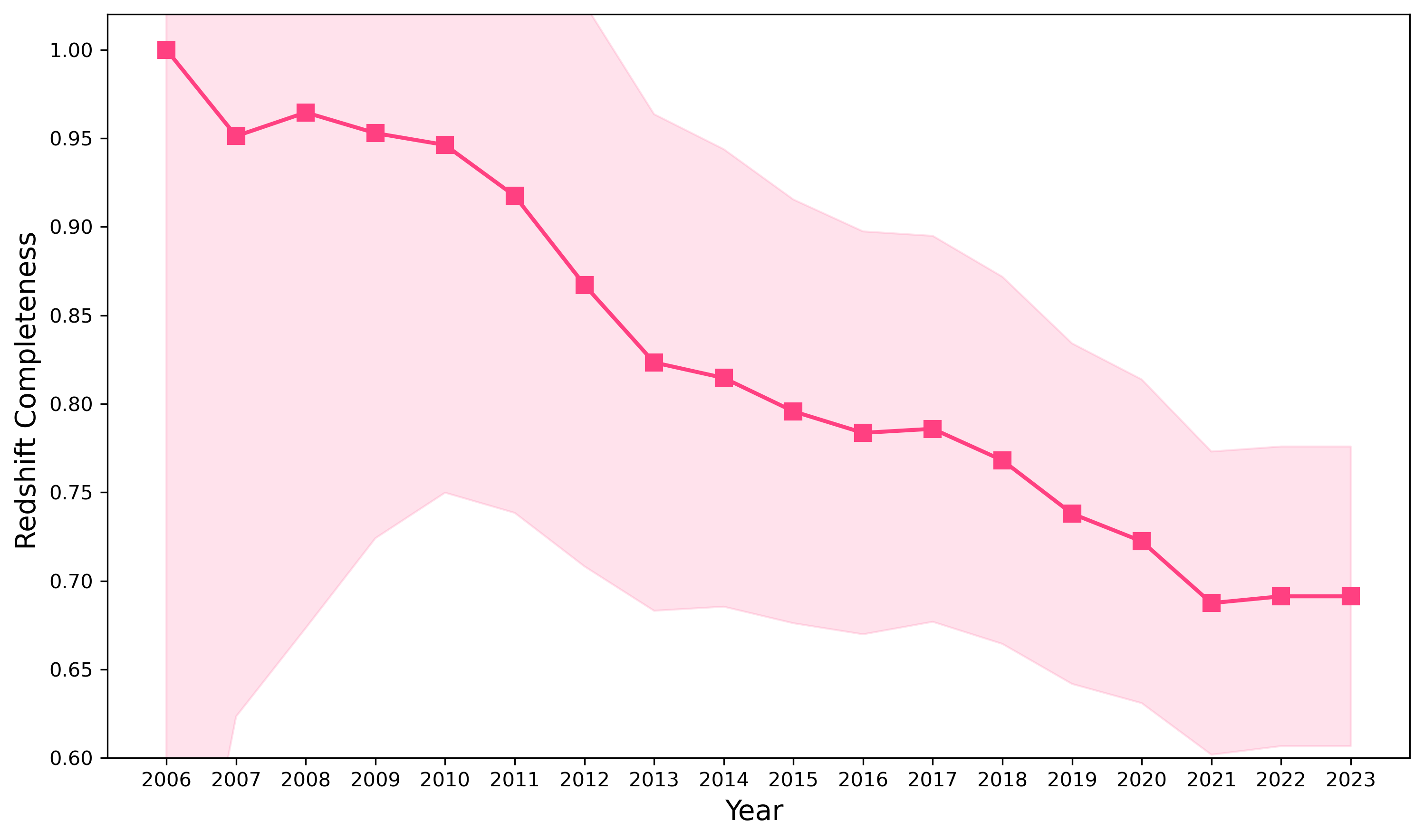}
\caption{
Redshift completeness as a function of observation time for Swift GRB samples selected according to the criteria defined by \protect\cite{2012ApJ...749...68S}. The shaded area indicates Poisson errors.
\label{fig:zC}}
\end{figure}

Applying these selection criteria to Swift data up to 2011, ~\cite{2012ApJ...749...68S} identified 58 LGRBs, of which 52 had measured redshifts, resulting in a redshift completeness of 90\%. Later, ~\cite{2016A&A...587A..40P} updated the dataset through July 2014 using the same criteria, identifying 99 LGRBs with 82 having measured redshifts, reducing the redshift completeness to 82\%.
We applied the aforementioned criteria to Swift LGRBs detected from 2004 to 2023, identifying a total of 160 LGRBs with a redshift completeness reduced to 68\%. Figure~\ref{fig:zC} illustrates the variation in redshift completeness of GRB samples selected according to Salvaterra’s criteria over the years. Interestingly, the redshift completeness shows a declining trend over time, possibly due to a decreasing fraction of GRBs with well-localized follow-up observations.

As shown in Figure~\ref{fig:zC}, the   redshift completeness declines significantly over time upon applying the Salvaterra criterion. This observation raises concerns about the suitability of the Salvaterra criteria for current Swift data, prompting us to seek  new selection criteria. Similar to the effect of  the peak flux threshold   on the redshift completeness~\citep{2023ApJ...958...37D}, other selection criteria, such as XRT error and Galactic extinction, can also significantly affect the redshift completeness. However, overly stringent criteria can lead to a rapid decline in the sample size. In extreme cases, redshift completeness can approach 100\%, but the sample size may reduce to fewer than ten events, which is clearly undesirable. Therefore, when determining the selection criteria, it is crucial to balance redshift completeness and sample size. We define the evaluation metric for the selection criteria as $F = N \times C^3$, where $N$ is the ratio of the number of samples after filtering to the total number of Swift bursts, and $C$ is the redshift completeness after filtering. We maintain the condition of peak flux $P \geq 2.6 \, \text{ph} \, \text{cm}^{-2} \, \text{s}^{-1}$ and utilize XRT error, Galactic extinction~($A_V$), GRB declination~($\delta$), and Sun-to-field distance~($\theta_{\text{Sun}}$) as free parameters to search for the parameters that maximize the corresponding F-value. The optimal parameters identified are (1) XRT error $<$ 4.59$^\circ$, (2) $A_V < 0.63$, (3) $\lvert \delta \rvert < 75.37^\circ$, and (4) $\lvert \theta_{\text{Sun}} \rvert > 9.87^\circ$. The resulting sample comprises 280 GRBs, among which 167 have measured redshift, exhibiting a redshift completeness of 60\%.
 In comparison to the sample derived from the direct application of the Salvaterra criteria, the redshift completeness decreases by 8\%, while the sample size increases by 75\%, representing a reasonable trade-off.

Table~\ref{table2} lists 280 Swift LGRBs that meet our selection criteria, which are categorized into three distinct groups with $z \geq 2$, $z < 2$, and no redshift measurements. Of these, 59 LGRBs are classified into the high-redshift group~($z \geq 2$). It was found that the observed event rate of LGRBs with $z \geq 2$ shows a trend closely matching the SFR~\citep{2015ApJS..218...13Y,2017ApJ...850..161T,2024ApJ...963L..12P}, supporting the hypothesis that they are predominantly collapsars. In this work, we constructed the LF for high-redshift LGRBs with $z \geq 2$ and extend  to  the lower redshift.

\section{ANALYSIS METHOD}
 \label{sec:ANALYSIS METHOD}

\begin{figure}
	\includegraphics[width=\columnwidth]{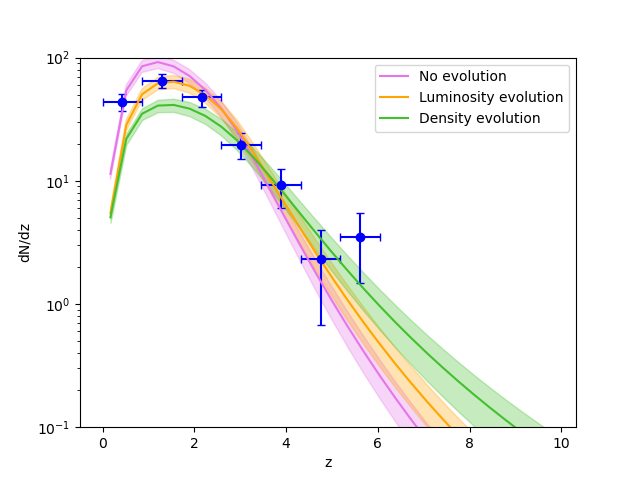}
\caption{
 The differential redshift distributions of three theoretical  models for collapsar GRBs~(color curves). shaded areas indicate the 1 $\sigma$ confidence intervals for each model.  The blue data points represent the observed redshift distribution of 167 LGRBs with $P \geq 2.6$ ph cm$^{-2}$ s$^{-1}$.
 }

\label{fig:dNdz}
\end{figure}

\begin{figure*}
    \centering
    \hspace{-0.38cm}
    \includegraphics[width=0.835\textwidth]{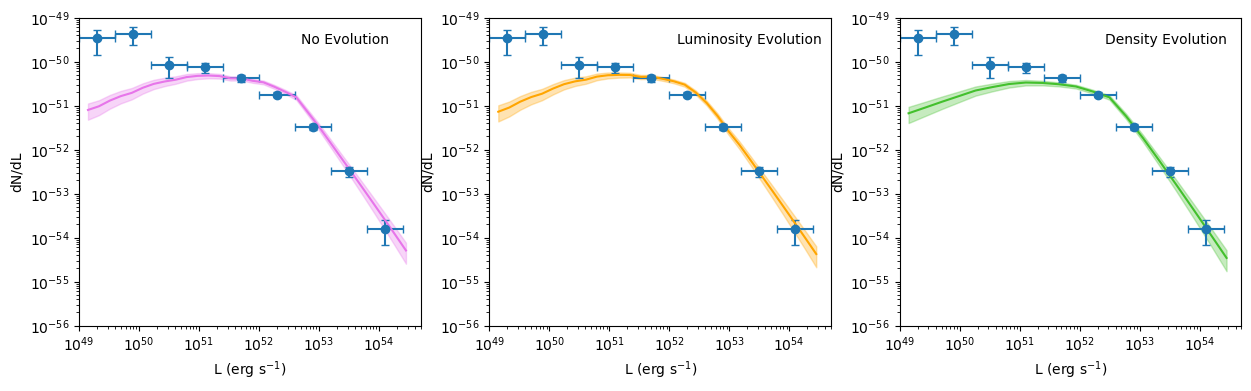}
    \caption{The blue data points show the luminosity distributions of 167  LGRBs with $P \geq 2.6 \ \text{ph cm}^{-2} \ \text{s}^{-1}$. The solid lines  and the shaded regions stand for the luminosity distributions and 1$\sigma$ scatters of collapsar GRBs as predicted by different models. }
    \label{fig:dNdL}
\end{figure*}

\begin{figure*}
    \centering
    \includegraphics[width=\textwidth]{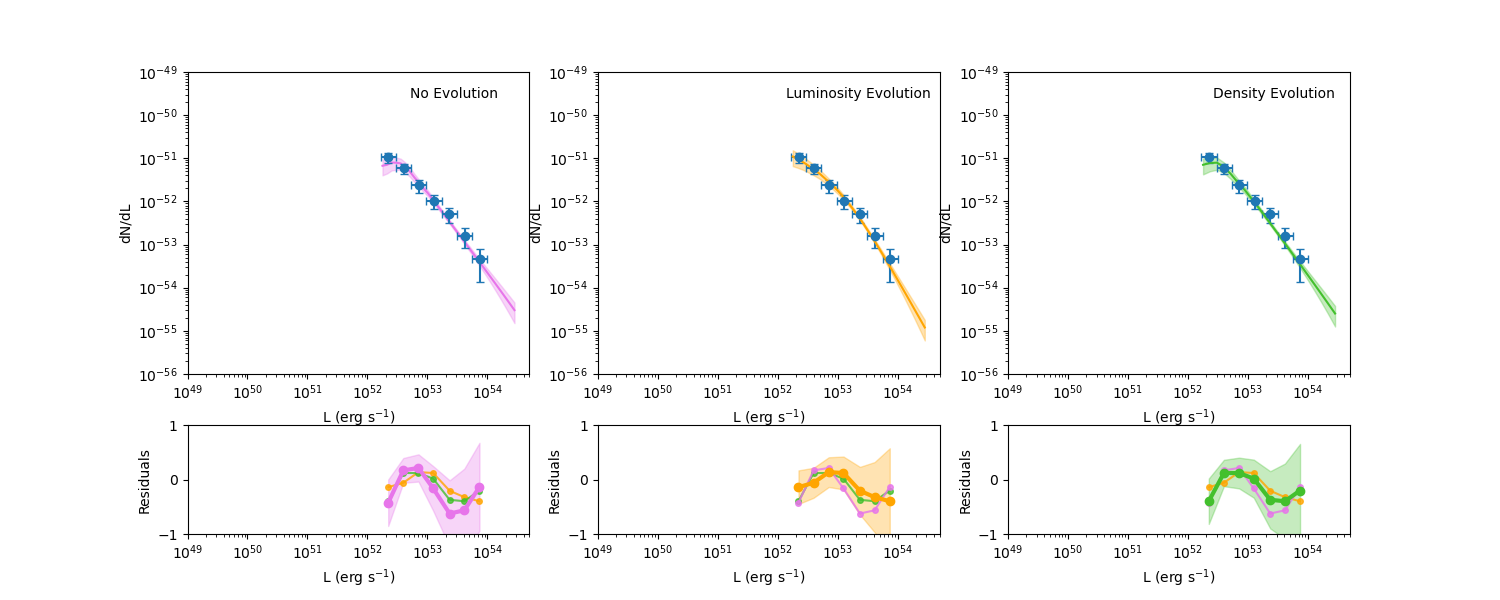}
   
    \caption{
    Luminosity distributions of 59 high-redshift ($z \geq 2$) LGRBs with $P \geq 2.6 \ \text{ph cm}^{-2} \ \text{s}^{-1}$ (solid circles, top row). The top row presents the best-fitting models(solid lines) for three different evolutionary scenarios: no evolution (left), luminosity evolution (middle), and density evolution (right). The shaded regions represent the 1$\sigma$ confidence intervals. The bottom row displays the residuals corresponding to each model. For each model's residual plot, the residuals for all three models are plotted, with the residuals for the corresponding model shown in bold. 
    The color of the curves in the residual plots corresponds to the color of the respective models in the top row.}
    \label{fig:dNdL_zLT2}
\end{figure*}

We adopted the maximum likelihood method  to optimize the free parameters of the model~\citep{1983ApJ26935M,2019MNRAS.490..758Q,2006ApJ...643...81N,2009ApJ...699..603A,2012ApJ...751..108A,2010ApJ...720..435A,2014MNRAS.441.1760Z,2016MNRAS.462.3094Z,2019MNRAS.488.4607L,2021MNRAS.508...52L,2022ApJ...938..129L}, i.e.,

\begin{equation}
    \mathcal{L}=e^{(-N_{exp})} \prod_{i=1}^{N_{obs}} \Phi(L_{i}, z_{i}, t_{i})  ,
    \label{eq:L}
\end{equation} 
where $N_{exp}$ and $N_{obs}$ represents the expected and observed numbers of LGRBs, respectively. The term $\Phi(L, z, t)$ denotes the observed rate of GRBs per unit time, within the ranges of  $z$ to $z + dz$, $L$ to $L + dL$. The specific form of $\Phi(L, z, t)$ is given by

\begin{equation}
\begin{split}
\Phi(L, z, t) = \frac{d^{3} N}{dtdzdL} = \frac{d^{3}N}{dtdVdL} \times \frac{dV}{dz} ,
\end{split}
\label{eq:Phi}
\end{equation}
Thus
\begin{equation}
\begin{split}
\Phi(L, z, t) = \frac{\Delta \Omega}{4\pi}  \theta(P)  \frac{\psi(z)}{(1+z)}  \phi(L, z) \times \frac{dV}{dz} ,
\end{split}
\label{eq:Phi2}
\end{equation}
in which $\Delta \Omega = 1.33 \, \text{sr}$ represents the field of view~(FOV)  of  Swift/BAT. $\theta(P)$ indicates the detection efficiency that can be expressed as $\theta(P) \equiv \theta_{\gamma}(P) \theta_{z}(P)$, where $\theta_{\gamma}(P)$ represents the probability of detecting a burst for a given peak flux $P$. Since our sample is selected with lower limit of  $P \geq 2.6 \, \text{ph} \, \text{cm}^{-2} \, \text{s}^{-1}$, the selection effect is negligible, enabling us to assume $\theta_{\gamma}(P)=1$ ~\citep{2021MNRAS.508...52L}. $\theta_{z}(P)$ corresponds to the probability of measuring the redshift for a given peak flux $P$. For relatively higher redshift-complete samples~\citep{2012ApJ...749...68S,2016A&A...587A..40P} 
, the term $\theta_{z}(P)$ was usually  approximated to be 1 ~\citep{2019MNRAS.488.4607L}, which is reasonable given their redshift completeness exceeding 80\%. However, our sample has a redshift completeness of 60\%, so we fit with an empirical function to  the relationship between $\theta_{z}(P)$ and $P$, as $\theta_{z}(P) = {(1 + (1.28 \pm 0.21) \times (0.95 \pm 0.01)^P)^{-1}}$.
Under the collapsar  hypothesis,  the GRB event rate, $\psi(z)$, is expected to  intrinsically follow  the SFR, $\psi_{\star}(z)$, namely $\psi(z) = \eta \psi_{\star}(z)$, where $\eta$ denotes the efficiency of GRB formation. Following~\cite{2006ApJ...651..142H} and~\cite{2008MNRAS.388.1487L}, the SFR, $\psi_{\star}(z)$,  can be described by as:

\begin{equation}
 \psi_{\star}(z) = \frac{0.0157 + 0.118z}{1 + (z/3.23)^{4.66}} \quad [M_{\odot} \, \text{yr}^{-1} \, \text{Mpc}^{-3}] ,
\end{equation}
The function $\phi(L, z)$ represents the normalized GRB LF, which may evolve with redshift depending on the  detailed model below. Here, we adopt a generic form of a broken power-law as:

\begin{equation}
\phi(L, z) = \frac{A}{\ln(10) L} \left\{
\begin{array}{ll}
\left(\frac{L}{L_{c}(z)}\right)^{a}, & L \leq L_{c}(z) \\
\\
\left(\frac{L}{L_{c}(z)}\right)^{b}, & L > L_{c}(z)
\end{array}
\right.,
\end{equation}
where $A$ is the normalization constant, $a$ and $b$ are the power-law indices corresponding to luminosities below and above the break luminosity $L_{c}(z)$, respectively.


The expected number of GRBs can be expressed as

\begin{equation}
\begin{split}
    N_{\text{exp}} = \frac{\Delta \Omega T}{4 \pi} \int_{z_{\text{min}}}^{z_{\text{max}}} \int_{\max[L_{\text{min}}, L_{\lim}(z)]}^{L_{\text{max}}} \theta(P(L, z)) \frac{\psi(z)}{1+z} \\ 
    \times \phi(L, z) \, dL \, \frac{dV}{dz} \, dz,
\end{split}
\label{eq:Nexp}
\end{equation}
The luminosity threshold in Eq.~(\ref{eq:Nexp}) can be computed by:

\begin{equation}
    L_{\lim}(z) = 4 \pi D_{L}^{2}(z) P_{\lim} \frac{\int_{1/(1+z) \, \text{keV}}^{10^4/(1+z) \, \text{keV}} E N(E) \, dE}{\int_{15 \, \text{keV}}^{150 \, \text{keV}} N(E) \, dE},
\end{equation}
where $P_{\lim}$ represents the peak flux threshold of our sample, that is $P_{\lim} = 2.6 \, \text{ph} \, \text{cm}^{-2} \, \text{s}^{-1}$.  $N(E)$ describes the GRB photon spectrum, that is generally modeled by a  Band function~\citep{1993ApJ...413..281B, 2006ApJS..166..298K}, with low~(high) energy  spectral indice of $-1$~($-2.3$). 

To account for (i) the potential biases in the relationship between GRB event rates and the SFR, and (ii) how GRB LF evolves with redshift, we introduce an additional evolutionary factor of  $(1 + z)^\delta$ into the Eq~(~\ref{eq:Nexp}), in which $\delta$ is a free parameter. We investigate three distinct models: (1) the GRB event rate follows the SFR directly, i.e., $\psi(z) = \eta\psi_{\star}(z)$, with a non-evolving LF, i.e., $L_{ci}(z) = L_{ci, 0}$ ~(constant); (2) the GRB event rate is proportional to the SFR, but the LF's break luminosity evolves with redshift, as $L_{ci}(z) = L_{ci, 0}(1 + z)^\delta$; and~(3) the GRB event rate includes an additional evolutionary factor, say $\psi(z) = \eta\psi_{\star}(z)(1 + z)^\delta$, while  the LF is assumed to be independent of redshift. 

Once the optimal parameters for each model are determined, we substitute them into Eq.~(\ref{eq:Nexp}), and set the redshift integration limits from $z_{\text{min}}=0$ to $z_{\text{max}}=2$. This allows us to compute the expected number of collapsar GRBs between redshifts 0 and 2 of any one of the above three models.

\section{RESULTS} \label{sec:RESULTS}

We applied the MCMC method to optimize the model's free parameters~\citep{2010CAMCS...5...65G,2013PASP..125..306F}, specifically the GRB LF, the evolution parameter, and the formation efficiency $\eta$. Afterward, we computed the corresponding maximum likelihood estimation, Akaike Information Criterion~(AIC), and Bayesian Information Criterion~(BIC). The results are summarized in Table~\ref{table1}. Figures ~\ref{fig:dNdz} and~\ref{fig:dNdL} illustrate the redshift and luminosity distributions of 167 GRBs with measured redshift and~$P \geq 2.6 \,\mathrm{ph}\ \mathrm{cm}^{-2} \,\mathrm{s}^{-1}$ in our sample. The solid lines represent the extrapolated redshift and luminosity distributions  by extending the LF, which was fitted to the high-redshift sample but down to $z=0$. In other words, in Figure \ref{fig:dNdz}, the data points represent the redshift distribution of the LGRBs, while the solid lines show the redshift distributions of the collapsar GRBs derived from three different model. Similarly, in Figure \ref{fig:dNdL}, the data points represent the luminosity distribution of the LGRBs, and the solid lines correspond to the luminosity distribution of the collapsar GRBs   as derived in this study.

 In Figure \ref{fig:dNdz}, the redshift distribution of collapsar GRBs obtained from the no-evolution model~(the pink solid line) is noticeably higher at low redshifts than the redshift distribution of LGRBs~(the blue data points). It is worth noting that similar results have been observed in previous studies of the LGRB LFs ~\citep{2012ApJ...749...68S,2019MNRAS.488.4607L,2021MNRAS.508...52L}, where the no-evolution model also shows a significantly higher redshift distribution  at low redshifts compared to the observed distribution. This discrepancy is attributed to the fact that an evolution term is essential for the LF of LGRBs, and the no-evolution model fails to adequately fit the data.


Within the luminosity evolution model framework, the GRB event rate exhibits more close to the SFR. However, the characteristic break luminosity of the GRB LF  evolves with the redshift, namely $L_c(z) = L_{c, 0}(1 + z)^{\delta}$.
Utilizing a BPL LF, our analysis demonstrates notable luminosity evolution, yielding a value of $\delta = 1.87^{+0.27}_{-0.31}$. This result is consistent with previous parametric studies~(e.g., ~\citeay{2012ApJ...749...68S}, ~\citeay{2019MNRAS.488.4607L}). Additionally, this value is roughly consistent with estimates obtained using non-parametric methods~(e.g.,~\citeay{2002ApJ...574..554L},~\citeay{2015ApJ...806...44P},~\citeay{2015ApJS..218...13Y},~\citeay{2016A&A...587A..40P}, ~\citeauthor{2022MNRAS.513.1078D} ~\citeyear{2022MNRAS.513.1078D},~\citeyear{2023ApJ...958...37D}). Based on the optimal parameters of the luminosity evolution model, we calculate the expected number of collapsar GRBs within the  range of 0<z<2. By adjusting the limits of integration in Eq.~(\ref{eq:Nexp}) to this range
, the luminosity evolution model predicts that the number of collapsar GRBs with $z < 2$ and $P \geq 2.6 \, \text{ph} \, \text{cm}^{-2} \, \text{s}^{-1}$ is 72.67, representing 67.29\% of the observed LGRBs. This indicates that approximately 32.71\% of low-redshift LGRBs are likely not associated with collapsars.

The density evolution model assumes that the break luminosity in the GRB LF remains constant, while the GRB event rate behaves a redshift-dependent feature of  $\psi(z) = \eta \psi_{\star}(z)(1 + z)^{\delta}.$ Using the BPL LF again, one can constrain the density evolution to be  $\delta = 1.10^{+0.21}_{-0.20}$.
We apply Eq.~(\ref{eq:Nexp}) to  estimate the expected number of collapsar GRBs within the redshift range of $0 < z < 2$. As a result, the density evolution model predicts that the number of collapsar GRBs with $z < 2$ and $P \geq 2.6 \, \text{ph} \, \text{cm}^{-2} \, \text{s}^{-1}$ is 57.28, corresponding to  53.04\% of the observed LGRBs. This indicates that approximately 47\% of low-redshift LGRBs may non-collapsars. 


It can be seen from Figure~\ref{fig:dNdz}  that at low redshifts, the redshift distribution of the collapsar LF obtained from the luminosity evolution~(yellow solid line) and density evolution~(green solid line) is significantly lower than the  redshift distribution of observed LGRBs~(the blue data points). Our findings are in good agreement with the results of ~\cite{2024ApJ...963L..12P}, who found that approximately 60\% of LGRBs with $z < 2$ are not collapsar events. 

From the perspective of luminosity distribution ~(Fig.~\ref{fig:dNdL}), the luminosity distribution of collapsar GRBs (solid line) is significantly lower than that of LGRBs (the blue data points) in the low-luminosity part. This is consistent with the findings of ~\cite{2023ApJ...958...37D}, which show that the event rate of high-luminosity LGRBs follows the SFR well, while the event rate of low-luminosity LGRBs deviates significantly from the SFR, providing mutual confirmation.

To demonstrate that our model fits the sample well, we present in Figure~\ref{fig:dNdL_zLT2} the luminosity distributions ~(solid circles) of 59 high-redshift ~($z \geq 2$) LGRBs with $P \geq 2.6 \ \text{ph cm}^{-2} \ \text{s}^{-1}$, which is the high-redshift collapsar sample used for our fit, as well as the luminosity distributions for $z > 2$ from the three models. In the second row of Figure \ref{fig:dNdL_zLT2}, we plot the residuals for each model. The residuals are defined as $Residuals = \frac{dN/dL_{\text{theory}} - dN/dL_{\text{obs}}}{dN/dL_{\text{theory}}}$. For each model's residual plot, the residuals for all three models are plotted, with the residuals for the corresponding model shown in bold. The color of the curves in the residual plots corresponds to the color of the respective models in the top row. From the perspective of luminosity distribution, all three models provide a good fit to the sample. From the residual plot, the no-evolution model slightly underperforms compared to the other two models.

It should be emphasized that we are not directly fitting the luminosity and redshift distributions, but instead obtaining the best-fit parameters via the maximum likelihood method using Equation~(\ref{eq:L}).  We also use AIC and BIC to evaluate the relative performance of the models.

After obtaining the AIC for each model, we can use the Akaike weight $exp(-AIC_i/2)$ to judge the confidence level of one model relative to another. The relative probability of the $i^{th}$ model being preferred is defined specifically as :
\begin{equation}
    P(M_i) = \frac{e^{-AIC_i / 2}}{e^{-AIC_1 / 2} + e^{-AIC_2 / 2}},
    \label{eq:aIC}
\end{equation}
Based on the AIC, the probability that the no-evolution model is correct compared to the luminosity evolution model is only 0.001. Furthermore, the difference in BIC values between the no-evolution model and the luminosity evolution model is 8.12, which is sufficient to rule out the no-evolution model.

\section{Conclusion and Discussion} \label{sec:cite}
We re-assessed the selection criteria for high redshift-completeness GRB samples and applied the updated  criteria to identify a sample of 280 GRBs with peak flux $P \geq 2.6 \, \text{ph} \, \text{cm}^{-2} \, \text{s}^{-1}$, which is 2.8 times larger than the BAT6ext sample~\citep{2016A&A...587A..40P}. Using GRBs with $z \geq 2$, we constructed the LF and evaluated three evolutionary scenarios including no evolution, luminosity evolution, and density evolution. Our analysis shows that the no evolution scenario can be ruled out, while both the luminosity evolution and the  density evolution models fit the observed data well.
The luminosity evolution model predicts that the number of collapsar GRBs with $z < 2$ and $P \geq 2.6 \, \text{ph} \, \text{cm}^{-2} \, \text{s}^{-1}$ is 72.67, accounting for 67.29\% of the observed LGRBs. Similarly, the density evolution model predicts that the number of collapsar GRBs under the same conditions is 57.28, representing 53.04\% of the observed LGRBs. Our results confirm that the majority of low-redshift~(z<2) LGRBs are produced from the collapsar rather than the non-collapsar progenitors. Our study suggests that LGRBs originating from sources other than massive star collapses may not be rare occurrences. In fact, for redshifts below 2, the fraction could be as high as 33\% under the luminosity evolution model, and up to 47\% under the density evolution model.

As of 2009,~\cite{2009ApJ...703.1696Z} identified three LGRBs~(050724, 060614 and 061006)  as the Type I burst, of which their progenitors were likely mergers of two compact stars. More recently, further evidence has emerged for additional  LGRBs with strong indications of compact star mergers as progenitors, such as GRB 211211A~\citep{2022Natur.612..223R} and GRB 230307A~\citep{2024Natur.626..737L}, as well as GRB 170228A, which exhibits similar characteristics~\citep{2024arXiv240702376W}. 

Regarding the luminosity distributions, how to choose the function is still controversial.   For example,~\cite{2015ApJ...812...33S} identified that the LF of LGRBs is best described by a TPL. Similarly,~\cite{2021MNRAS.508...52L} using a larger Swift sample, reported consistent findings. It is noteworthy that a TPL is an uncommon 
form for LFs. In contrast,~\cite{2015ApJ...812...33S} found that the LFs of other extragalactic high-energy transients, such as SGRBs, supernova shock breakouts~(SBOs), and tidal disruption events~(TDEs), typically follow a single power law rather than a TPL. A plausible explanation for the TPL in the LGRB LF is the inclusion of a subset of non-collapsar bursts. These non-collapsar bursts may exhibit a BPL LF with a lower luminosity, while collapsar bursts have their own BPL LF with higher luminosity. The combination of these two components could cause the LGRB LF to appear as a TPL.~\cite{2023ApJ...958...37D} classified LGRBs into high- and low-luminosity groups and calculated their event rates separately. Their analysis revealed that the event rate of high-luminosity GRBs closely follows the SFR, whereas the event rate of low-luminosity GRBs deviates significantly, further supporting this argument.

The nature of non-collapsar LGRBs at low redshift still remains an open question. One plausible hypothesis suggests that non-collapsar LGRBs may originate from compact binary mergers, such as black hole–black hole mergers~\citep{2022A&A...657L...8B} or white dwarf–black hole mergers~\citep{2024arXiv240812654L}. However, several non-compact star merger scenarios have also been proposed, such as helium mergers~(or similar type ) systems~\citep{2019ApJ...871..118L}, and the collapse of massive stars in interacting binary systems~\citep{2022ApJ...928..104L}. If the progenitors of these LGRBs are stars, their event rate should correlate with the SFR. In contrast, if they arise from compact binary mergers, their event rate would be associated with the delayed SFR~\citep{2018MNRAS.477.4275P,du2024}. Constructing LFs based on both  SFR and  delayed SFR, and fitting the excess in the dN/dZ and dN/dL distributions seen in Figures~\ref{fig:dNdz} and~\ref{fig:dNdL} compared to the predictions for collapsar GRBs, may help identify the origins of these non-collapsar GRBs.

\begin{table*}
\centering
\renewcommand{\arraystretch}{1.5} 
\caption{Parameters of the best fit with different models. }
\begin{tabular}{ccccccccc}
\toprule
Model & Evolution parameter &$\eta$ &  $a$ & $b$ & log $L_c$  & ln $\mathcal{L}$ & AIC&BIC \\
 &  &  ($10^{-8} M^{-1} _{\odot}$) &  & & (erg s$^{-1}$)&   & & \\ \midrule

No evolution &… &$8.78^{+1.17}_{-1.19}$ & $-0.08^{+0.02}_{-0.02}$ & $-1.07^{+0.18}_{-0.17}$ & $52.57^{+0.13}_{-0.14}$ & -82.85 & 173.70&182.01 \\
Luminosity evolution & $\delta = 1.87^{+0.27}_{-0.31}$ &$6.69^{+0.95}_{-0.93}$ & $-0.29^{+0.05}_{-0.05}$ & $-1.59^{+0.19}_{-0.18}$ & $52.20^{+0.11}_{-0.12}$ & -76.75 & 161.5&173.89 \\
Density evolution & $\delta = 1.10^{+0.21}_{-0.20}$ &$ 4.23 ^{+0.83}_{-0.85}$&$-0.21^{+0.03}_{-0.03}$ & $-1.11^{+0.21}_{-0.19}$ & $52.54^{+0.10}_{-0.11}$ & -76.89 & 161.78&174.17 \\ 

\bottomrule
\end{tabular}
\begin{minipage}{\textwidth}
\footnotesize
\textbf{Note.} The parameter values were determined as the medians of the best-fitting parameters from the Monte Carlo sample. The errors represent the 68\% containment regions around the median values.
\end{minipage}
\label{table1}
\end{table*}

\begin{table*}
\centering
\caption{List of 280 Swift GRBs that meet our selection criteria.\label{table2}}

\begin{tabular}{c c c c c c c c c c c c}
\hline
\hline
\multicolumn{12}{c}{GRBs with z $\geq 2$  } \\ 
\hline
GRB & Ref & GRB & Ref & GRB & Ref & GRB & Ref & GRB & Ref & GRB & Ref \\

\hline																																			
050603	&	3			&	070714A	&	3			&	090201	&	1	,	3	&	120811C	&	15			&	150206A	&	1	,	4	&	190719C	&	1	,	11	\\
050922C	&	1	,	3	&	080413A	&	2	,	3	&	090715B	&	16			&	121209A	&	1	,	3	&	150403A	&	1	,	4	&	191004B	&	2	,	12	\\
051008	&	1	,	4	&	080603B	&	14			&	090812	&	16			&	130427B	&	1	,	3	&	150424A	&	1	,	2	&	210112A	&	1	,	2	\\
051109A	&	13			&	080607	&	14			&	100704A	&	2	,	3	&	130505A	&	1	,	4	&	151021A	&	1	,	5	&	220101A	&	1	,	2	\\
060206	&	1	,	4	&	080721	&	14			&	100728B	&	1	,	3	&	130514A	&	1	,	3	&	161014A	&	1	,	4	&	220521A	&	1	,	2	\\
060210	&	15			&	080804	&	15			&	110205A	&	1	,	3	&	130606A	&	1	,	3	&	161017A	&	1	,	6	&	230818A	&	1	,	2	\\
060927	&	13			&	081121	&	14			&	110709B	&	1	,	3	&	140206A	&	16	,	17	&	170202A	&	1	,	7	&	231210B	&	1	,	2	\\
061222A	&	2	,	3	&	081203A	&	2	,	3	&	110731A	&	1	,	3	&	140419A	&	1	,	3	&	170705A	&	1	,	8	&	221226B	&	1	,	2	\\
070328	&	1	,	4	&	081221	&	15			&	111008A	&	1	,	3	&	140629A	&	1	,	3	&	180325A	&	1	,	9	&	210411C	&	1	,	2	\\
070521	&	15			&	081222	&	14			&	120802A	&	1	,	3	&	140703A	&	1	,	3	&	181020A	&	1	,	10	&		&				\\
																																			
\hline																																			

\multicolumn{12}{c}{GRBs with z $< 2$  } \\ 
\hline
050219A  & 050318  & 050416A & 050525A & 050802  & 051111  & 060306  & 060418  & 060505  & 060614  & 060814  & 060908  \\

060912A  & 061007  & 061021  & 061121  & 061126  & 070306  & 070714B & 071112C & 071117  & 080319B & 080319C & 080411  \\

080413B  & 080430  & 080602  & 080605  & 080916A & 081007  & 090102  & 090424  & 090709A & 090926B & 091018  & 091020  \\

091127   & 091208B & 100615A & 100621A & 100728A & 100816A & 110422A & 111228A & 120119A & 120326A & 120729A & 120907A \\

130420A  & 130427A & 130701A & 130831A & 130907A & 130925A & 131030A & 131105A & 140213A & 140506A & 140512A & 141220A \\

141221A  & 150301B & 150314A & 150323A & 151027A & 160131A & 160410A & 160425A & 160703A & 160804A & 161001A & 161117A \\

161129A  & 161219B & 170604A & 170903A & 180205A & 180314A & 180620B & 180720B & 181110A & 190106A & 190114C & 190324A \\

190829A  & 191019A & 191221B & 200528A & 200829A & 201024A & 201104B & 201216C & 201221D & 210104A & 210207B & 210210A \\
210217A  & 210410A & 210610B & 210619B & 210822A & 211211A & 211227A & 220427A & 230325A & 231111A & 231117A & 231118A \\
\hline
\multicolumn{12}{c}{GRBs without redshift  } \\ 
\hline
050124  & 050128  & 050219B & 050326  & 060105  & 080218B & 080229A & 080613B & 090929B & 091221  & 100522A & 100619A \\

101017A & 110102A & 110402A & 110420A & 110709A & 110915A & 120102A & 120116A & 120311A & 120703A & 121123A & 121125A \\

130502A & 130527A & 130609B & 130727A & 130803A & 130812A & 131128A & 140102A & 140215A & 140529A & 140619A & 140628A \\

140716A & 140817A & 141005A & 141017A & 141212B & 150222A & 150428A & 150430A & 150607A & 150616A & 150710B & 150711A \\

151114A & 160119A & 160216A & 160325A & 160424A & 160607A & 161117B & 170126A & 170208A & 170208B & 170317A & 170626A \\

170803A & 170822A & 170906A & 170906C & 171027A & 171120A & 180111A & 180331B & 180404B & 180418A & 180620A & 180626A \\

180706A & 180809B & 180925A & 181126A & 190103B & 190202A & 190511A & 190604B & 190613B & 190824A & 191004A & 191031C \\

191122A & 200122A & 200125A & 200131A & 200215A & 200303A & 200306A & 200416A & 200716C & 200901A & 200922A & 201013A \\

201209A & 210226A & 210305A & 210306A & 210514A & 210730A & 220118A & 220408A & 220714B & 220826A & 221016A & 221027B \\

221201A & 230228A & 230510A & 230606A & 231205B &         &         &         &         &         &         &         \\
\hline
\end{tabular}

\begin{minipage}{\textwidth}
\footnotesize
\noindent \textbf{Note:} 280 Swift GRB used in this study. The spectra and redshift references are from the following sources: 
(1) \url{https://www.mpe.mpg.de/~jcg/grbgen.html};
(2) \url{https://swift.gsfc.nasa.gov/archive/grbtable/};
 (3)~\cite{2020ApJ...893...77W};
 (4)~\cite{2007ApJ...671..656B}; 
 (5)~\cite{2015GCN.18433....1G};
 (6)~\cite{2016GCN.20082....1F};
 (7)~\cite{2017GCN.20604....1F};
 (8)~\cite{2017GCN.21297....1B};
 (9)~\cite{2018GCN.22546....1F};
 (10)~\cite{2018GCN.23363....1T};
 (11)~\cite{2019GCN.25130....1P};
 (12)~\cite{2019GCN.25974....1S};
 (13)~\cite{2008MNRAS.391..577A};
 (14)~\cite{2009A&A...508..173A};
 (15)~\cite{2019MNRAS.486L..46A};
 (16)~\cite{2016A&A...585A..68W};
 (17)~\cite{2020ApJ...893...46V};
 \end{minipage}

\end{table*}

\section*{Acknowledgements}

This work was supported in part by National Natural Science Foundation of China~(grant No. U2031118) .

\section*{Data Availability}
All data used in this paper are public.



\bibliographystyle{mnras}
\bibliography{example} 





\bsp	
\label{lastpage}
\end{document}